\documentclass[twoside,12pt]{article}  %%% ÕâÊÇ LaTeX ¸ñʽÎļþ±ØÐëµÄÖ¸Á²»¿ÉȱÉÙ
\usepackage{CJK}   %%%%% ʹÓÃÖÐÎÄ¡¢ÈÕÎÄ¡¢º«ÎÄʱµ÷Óô˺ê°ü
\usepackage{indentfirst}  %% ½Ú±êÌâºóµÚÒ»ÐжÎËõ½ø
\usepackage{bm}    %%%%%  ÅÅÓ¡Êýѧ·ûºÅºÚбÌåʱµ÷Óô˺ê°ü
\usepackage{graphicx}  %%%% ÎÄÖвåÈë eps ͼÐÎʱµ÷Óô˺ê°ü£¬²åͼ·½·¨ 1
\begin{document}    %% Îı¾Îļþ¿ªÊ¼£¬ÕâÊDZØÐëµÄÖ¸Áî
%-------------------  First Head  -----------------------------------------
\thispagestyle{empty} \vspace*{0.8cm}\hbox
to\textwidth{\vbox{\hfill\huge\sf \hfill}}
\par\noindent\rule[3mm]{\textwidth}{0.2pt}\hspace*{-\textwidth}\noindent
\rule[2.5mm]{\textwidth}{0.2pt}

%=================== Text begin here =============================================

\begin{center}
\LARGE\bf Quantum walks on two kinds of two-dimensional models$^{*}$   %% ÂÛÎÄÌâÄ¿
\end{center}

\footnotetext{\hspace*{-.45cm}\footnotesize $^\dag$Dan Li. E-mail: lidansusu007@163.com }

\begin{center}
\rm Dan Li$^{\rm a,b)\dagger}$, \ \ Michael Mc Gettrick$^{\rm b)}$, \ \ Wei-Wei Zhang $^{\rm a)}$, \ and  \ Ke-Jia Zhang $^{\rm a)}$
\end{center}

\begin{center}
\begin{footnotesize} \sl
${State\ Key\ Laboratory\ of\ Networking\ and\ Switching\ Technology}$, \\${Beijing\ University\ of\ Posts\ and\ Telecommunications,\ Beijing,\ 100876,\ China}^{\rm a)}$ \\   %%%% µØÖ· a)
${The\ De\ Brun\ Centre\ for\ Computational\ Algebra,\ School\ of\ Mathematics, Statistics\ and\ Applied\ Mathematics}$, \\${National\ University\ of\ Ireland,\ Galway}^{\rm b)}$ \\   %%%% µØÖ· b)

%%% ¸ü¶àµØÖ·ÒÀ´ÎÍùÏÂÑÓÐø
\end{footnotesize}
\end{center}

\begin{center}
\footnotesize (Received X XX XXXX; revised manuscript received X XX XXXX)
          %% (Received ÈÕ ÔÂ Äê; revised manuscript received ÈÕ ÔÂ Äê)
\end{center}

\vspace*{2mm}

\begin{center}
\begin{minipage}{15.5cm}
\parindent 20pt\footnotesize
In this paper, we numerically study quantum walks on two kinds of two-dimensional graphs: cylindrical strip and Mobius strip. The two kinds of graphs are typical two-dimensional topological graph. We study the crossing property of quantum walks on these two models. Also, we study its dependence on the initial state, size of the model. At the same time, we compare the quantum walk and classical walk on these two models to discuss the difference of quantum walk and classical walk.
%%%% ÂÛÎÄÕªÒª
\end{minipage}
\end{center}

\begin{center}
\begin{minipage}{15.5cm}
\begin{minipage}[t]{2.3cm}{\bf Keywords:}\end{minipage}
\begin{minipage}[t]{13.1cm}
%%%%% ¹Ø¼ü´Ê
Quantum walk, Cylindrical strip,Mobius strip
\end{minipage}\par\vglue8pt
{\bf PACS: 03.67.Ac, 03.67.Lx, 02.30.Nw}
%%% PACS ·ÖÀàÂë
%% ²éѯÍøÖ·£ºhttp://www.aip.org/pacs
\end{minipage}
\end{center}

\section{Introduction}
\label{sec:level1}

Walk is one of the most basic model, no matter classical walks \cite{402,403,404}, i.e. Markov chain or quantum walks. One walker can walk in different kinds of topological graphs, among which line and circle are the most well-studied topic. Quantum walk \cite{000,401,405} is a powerful model that provides a method to explore all possible paths in a parallel way due to the constructive quantum interference along the paths, no matter in the discrete or the continuous version. Recent studies of quantum walks have suggested that Quantum walks constitute a promising ingredient in the research of quantum algorithms, ranging from element distinctness \cite{001} to database searching \cite{002,003,004,005}, from constructing quantum Hash scheme \cite{006,007} to graph isomorphism testing \cite{008,009}.

Much attention has been paid to quantum walks on different kinds of graphs \cite{010,012,013,301,302,304,305,306,307,011}.  In \cite{010}, the authors studied the quantum walks on regular graphs, such as cartesian lattice, triangular lattices, cycles with diagonals, complete bipartite graphs and `glued trees' graph. In \cite{012}, Ashley Montanaro considers quantum walks on reversible directed graphs. The authors consider quantum walks on percolation lattices, in which edges or sites are randomly missing in \cite{013}. Recently, much attention has been attached on three-state quantum walks \cite{301,302,304,305,306,307}. Furthermore, some authors propose the experiment for continuous-time quantum walks of photons in the cylindrical waveguide array, the Mobius strip waveguide array and the twisted circular waveguide array \cite{011}.

In this paper, we consider discrete quantum walks on two kinds of models: cylindrical strip and Mobius strip. The idea of study these two kinds of graphs is from considering these two models as two two-dimensional universes: parallel universe and wormhole. The walk on the cylindrical strip is like a two-dimensional walk in two parallel universes, where each side of the cylindrical strip represents a universe. Also, the Mobius strip is like the wormhole in  two-dimensional universe, in which a walker can get to a remote position in a short time. Since we are three-dimensional creatures who cannot image the four-dimensional universal, so the research about the two-dimensional universal from the three-dimensional view will be much easier. Also, Mobius strip has been used in compact resonator with the resonance frequency which is half that of identically constructed linear coils \cite{021}, in inductionless resistor \cite{022}, in superconductors with high transition temperature \cite{023} and in other chemistry/nano technology.

The paper is structured as follows. In Sect. \ref{sec:level2}, we introduce the mathematical model of the quantum walk on cylindrical strip. In Sect. \ref{sec:level3}, we study properties of the quantum walk on cylindrical strip. In Sect. \ref{sec:level4}, we introduce the mathematical model of the quantum walk on Mobius strip. In Sect. \ref{sec:level5}, we study properties of the quantum walk on Mobius strip. A short conclusion is given in Sect. \ref{sec:level6}.

\section{Quantum Walk on Cylindrical Strip}
\label{sec:level2}

In this part, we introduce the discrete quantum walk on cylindrical strip. The parametric equation of a cylindrical strip, whose center circle has radius 1, is
\begin{equation}\label{E21}
\left\{ \begin{array}{c}
          x=cos(u);\\
          y=sin(u);\\
          z=v.
        \end{array}\right.
\end{equation}where $0\leq u\leq2\pi$, $-1\leq v \leq1$. Figure 1 shows the cylindrical strip model. The red line means the circumference of the strip, whose length is $\textit{NL}$, i.e. $\textit{NL}$ points on the center circle. The height is $N$. For this two-dimensional model, there are two sides of the cylindrical strip, outside surface and inside surface. If a walker on the cylindrical strip can walk on both surface, it needs to cross the boundaries. Boundary condition has great effects on the evolution of quantum walk on graphs. Some authors have demonstrated periodic boundary conditions \cite{201}, non-reversal boundary conditions \cite{202}, absorbing boundary conditions \cite{203}and moving boundary conditions \cite{204}. In this paper, we consider the walker can cross the boundary as it does not exist. Combine the topological property of cylindrical strip, it is actually a periodic boundary condition.
\begin{figure}[hbt]\label{F11}
  \center
  \includegraphics[width=10cm]{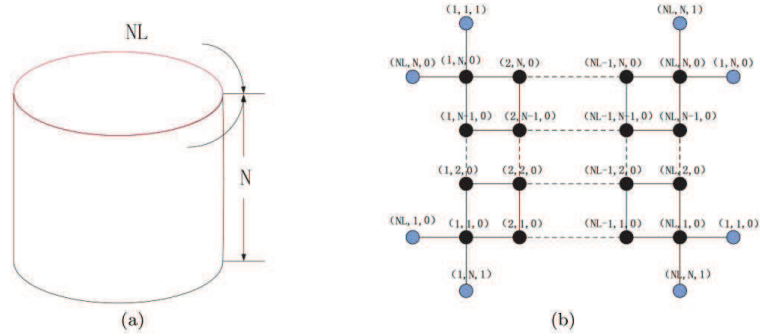}\\
  \center
  \caption{ Cylindrical strip model. Subgraph (b) demonstrates its corresponding lattice in two-dimension plane. The black vertices in the lattice are outside surface sites. }
\end{figure}

Next, we introduce the mathematical model of quantum walk on cylindrical strip. For the cylindrical strip, we use $d$ to indicate the side freedom of the strip. $d=0$ indicates the outside surface of the strip, while $d=1$ indicates the inside surface of the strip. We suppose every side of the strip is a $\textit{NL} \times \textit{N}$ lattice. Every lattice is the Hilbert space $ \mathcal{H}_p$ spanned by $|x,y\rangle$, where $x\in[0,\cdots, \textit{NL-1}]$, $y\in[0,\cdots, \textit{N-1}]$. The coin of the particle is a quantum system living in a four-dimensional Hilbert space $\mathcal{H}_c$ spanned by $\{00,01,10,11\}$. Therefore, the whole evolutive space for the walker is the Hilbert space $ \mathcal{H}$, whose basis states are ${|x,y,d,c_1,c_2\rangle}$.

The evolution of the whole system at each step of the walk can be described by the global unitary operator, denoted by $U$,
\begin{equation}\label{E21}
 U=S(\mathcal{I}\bigotimes C).
\end{equation}
C is the coin operation who acts on the $\mathcal{H}_c$, and S is the shift operation who acts on the $\mathcal{H}$.

What makes the quantum walk on the cylindrical strip special is the shift operation $S$. The walker moves to left or right, up or down for every step. The evolution of the walk is similar with that of the quantum walk on two-dimensional plane. However, when the walker goes up or down, it may cross  boundaries, and get into another side, i.e. the freedom of side $d$ will change. We should remind readers that the boundaries will not affect the evolution of the particle. Therefore, the movement on every direction is like moving on a circle with lazy action. Therefore, the shift operator S is
\begin{eqnarray}\label{E22}
  S & = & |x+1 (mod\ \textit{NL})\rangle\langle x|\otimes |y\rangle\langle y|\otimes|00\rangle_c \langle00|
 \nonumber\\
&& + |x\rangle\langle x|\otimes |y+1 (mod\ \textit{N})\rangle\langle y|\otimes|01\rangle_c \langle01|
\nonumber\\
&& +|x-1 (mod\ \textit{NL})\rangle\langle x|\otimes |y\rangle\langle y|\otimes|10\rangle_c \langle10|
\nonumber\\
&& +|x\rangle\langle x|\otimes |y-1 (mod\ \textit{N})\rangle\langle y|\otimes|11\rangle_c \langle11|.
\hspace{1mm}
\end{eqnarray}When $y+1\equiv 0(mod \ N)$ or $y-1\equiv \textit{N-1}(mod \ N)$, there is a X operator acts on the side freedom, ie. $d\equiv d+1(mod\ 2)$.

\section{Properties of the Quantum Walk on Cylindrical Strip}
\label{sec:level3}

In this part, we consider the properties of quantum walks on cylindrical strip. The choice of the initial state is important in studies of quantum walks, because interference features sensitively depend on the choice of the coin state. On the other hand, general properties do not depend on the choice of the coin state. So the coin state is not crucial provided here. We chose the initial position state $|0,0,0\rangle$, the initial coin state $[\frac{1}{2};\frac{1}{2};\frac{1}{2};\frac{1}{2}]$ and the coin operator $H$, which is the $4\times 4$ Hadamard operator
\begin{equation}\label{E23}
    H=\frac{1}{2}\left(
  \begin{array}{cccc}
    1 & 1 & 1 & 1 \\
    1 & -1 & 1 & -1 \\
    1 & 1 & -1 & -1 \\
    1 & -1 & -1 & 1 \\
  \end{array}
\right).
\end{equation}Mostly, we consider the cylindrical strip with $\textit{NL}=5$, and $\textit{N}=7$.

\subsection{general properties of probability distribution}
\label{sec:level31}

\begin{figure}[hbt]\label{F12}
  \center
  \includegraphics[width=10cm]{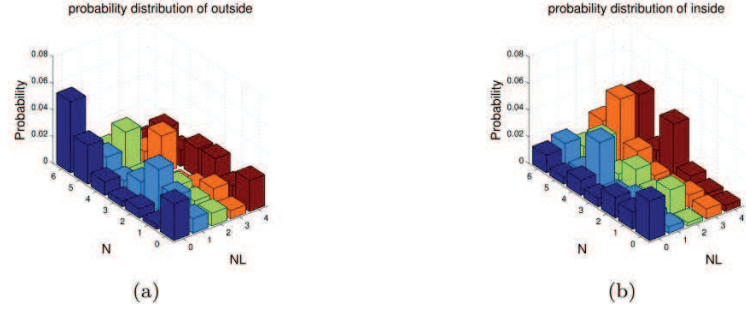}\\
  \center
  \caption{ The probability distribution of the quantum walk on cylindrical strip. The subgraph (a) shows the probability distribution of outside, and the subgraph (b) shows the probability distribution of inside.Time is 400.  }
\end{figure}

To begin with, we show the probability distribution of the quantum walk on the cylindrical strip in Fig. 2. The model with $\textit{NL}=5$, and $\textit{N}=7$ has  $2\times5\times7=70$ points. Even though each movement direction of the quantum walk is moving on a circle, because the movement cannot be seen as the tensor of two directions, the quantum walk on cylindrical strip is not the tensor of two quantum walks on a circle.

If $\textit{NL}$ and $N$ are not both odd, when the initial position state is not entangled state, the probability distribution has zeros at half positions. This property is same with graphs like circles and torus.  Because the evolution of the whole quantum system is unitary, the probability distribution of the walker will not converge, and keep fluctuating, which is different with classical walks.

\subsection{Crossing the boundary}
\label{sec:level32}

For quantum walks on cylindrical strip, the most important property is crossing the boundary and getting into the other side. Here we compare the crossing property of quantum walk with that of classical walk in Fig. 3. The classical walk has the same probability to move to the four adjacent points.

\begin{figure}[hbt]\label{F13}
  \center
  \includegraphics[width=10cm]{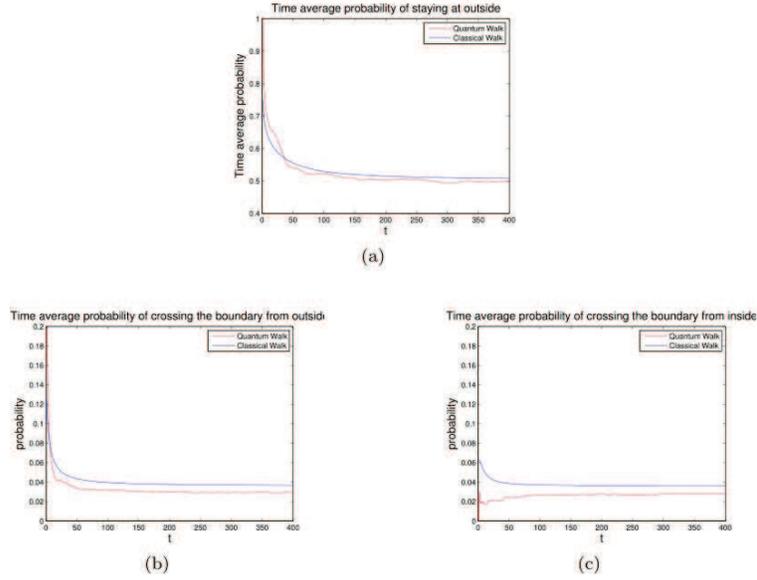}\\
  \center
  \caption{Subgraph (a) shows the probability of the particle staying at outside. Subgraph (b) shows the probability of the particle crossing the boundary from outside. Subgraph (c) shows the probability of the particle crossing the boundary from inside. }
\end{figure}

Because classical evolution has stable state, the classical walker on the cylindrical strip has 0.5 probability to stay on each side, which is according with our forecast. Also it has 0.0357 to cross the boundary from one side to another side. The probability to cross the boundary is 0.0714, which is twice of the probability to cross from outside. But the quantum walk is different from the classical walk. Because the quantum characteristic, the probability of crossing the boundary fluctuates wildly. Therefore, we consider the time average probability, which is defined as follows.
\begin{equation}\label{E24}
\widetilde{P}(T)=\frac{1}{T}\sum_{t=1}^{T}p(t)
\end{equation} Time-average probability of quantum walks will converge, so we can compare properties of different quantum walks easily .

It is obviously that quantum walk on cylindrical strip has same time average probability 0.5 to be at one side. But the quantum walk has lower time average probability to cross the boundary. It means that a quantum walker prefer staying at same side rather than crossing the boundary.

\subsection{Initial position state}
\label{sec:level33}

Then we want to study the effect of the initial state on the crossing property.  For most initial state, the probability of staying at each side or crossing the boundaries always fluctuates wildly. However, for one class of entangled position states $\alpha|x,y,0\rangle+\beta|x,y,1\rangle$, where $|\alpha|=|\beta|=\frac{1}{2}$, the situation changes.

\begin{figure}[hbt]\label{F14}
  \center
  \includegraphics[width=10cm]{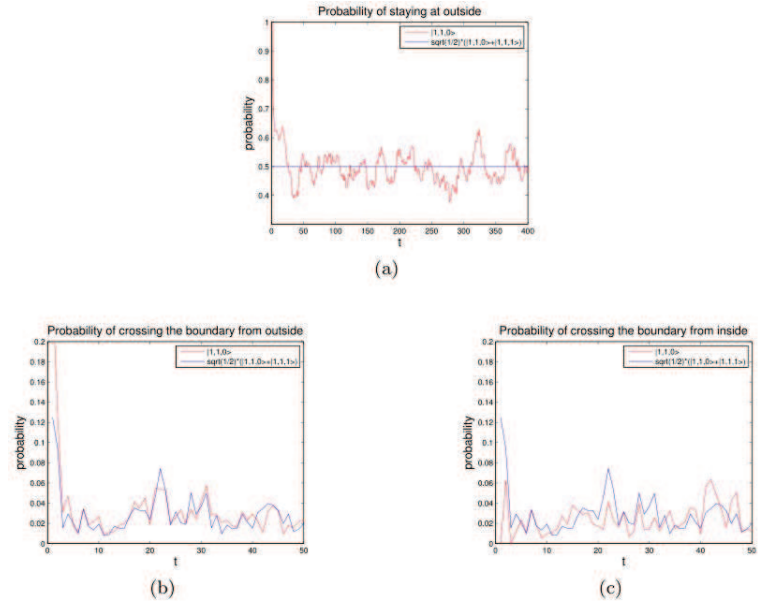}\\
  \center
  \caption{ The figures show the crossing property of quantum walks with initial position state $|1,1,0\rangle$ and $\frac{1}{\sqrt{2}}(|1,1,0\rangle+|1,1,1\rangle)$. Initial coin state is still $[\frac{1}{2};\frac{1}{2};\frac{1}{2};\frac{1}{2}]$. }
\end{figure}

For initial position state
\begin{equation}\label{E25}
\alpha|x,y,0\rangle+\beta|x,y,1\rangle,
\end{equation}
we suppose the finial state from $|x,y,0\rangle$ is $\psi_{\alpha}$, and that from $|x,y,1\rangle$ is $\psi_{\beta}$. It is obvious that $\psi_{\alpha}$ resembles $\psi_{\beta}$ except the freedom of side. Therefore,
\begin{equation}\label{E26}
\psi_{\alpha}(in\rightarrow out)=\psi_{\beta}(out\rightarrow in),
\end{equation}
\begin{equation}\label{E27}
\psi_{\beta}(in\rightarrow out)=\psi_{\alpha}(out\rightarrow in).
\end{equation}
As mentioned earlier, the movement in the Y-coordinates is on a even circle. So there is no inference between $\psi_{\alpha}$ and $\psi_{\beta}$. Given all this, we can get the conclusion that when $|\alpha|=|\beta|$, quantum walk with initial position state $\alpha|x,y,0\rangle+\beta|x,y,1\rangle$ always have the probability 0.5 to stay at each surface of the cylindrical strip. In Fig. 4, by using this class of states as initial position state, we can see the result that the probability of crossing the boundary from outside to inside equals to that of crossing the boundary from inside to outside at any time. So the probability of staying at outside or inside is $0.5$ forever. Most of all, this property holds for different size of cylindrical strip.

\subsection{Size of cylindrical strip}
\label{sec:level34}

In the above sections, we only consider the cylindrical strip with $\textit{NL}=5$ and $\textit{N}=7$. Now we want to research the effect of the size of cylindrical strip. We can image that with the size increasing, the crossing probability will decrease. Indeed it is. In the Fig. 5, we study the crossing property with $\textit{NL}$, $N$ changing. The initial position state is still $|1,1,0>$.
\begin{figure}[hbt]\label{F15}
  \center
  \includegraphics[width=10cm]{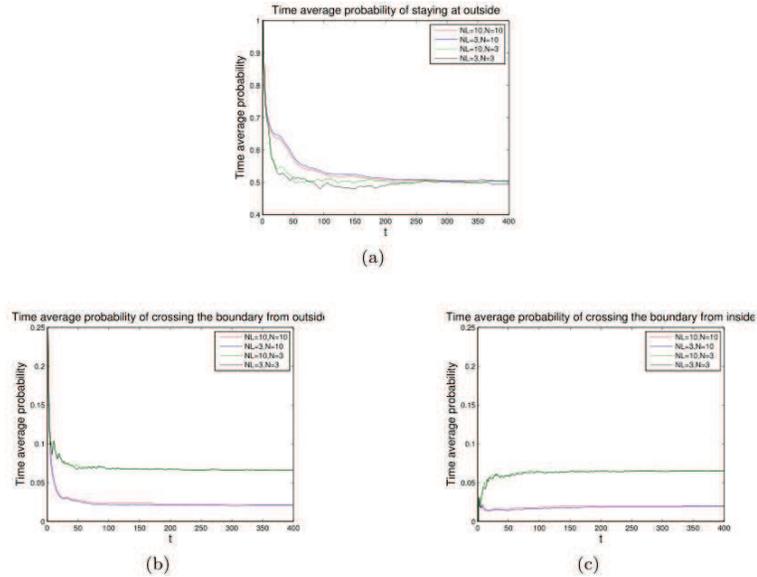}\\
  \center
  \caption{ The figures show the crossing property of quantum walks on cylindrical strip with $\textit{NL}=10, N=10$; $\textit{NL}=3, N=10$; $\textit{NL}=10, N=3$; $\textit{NL}=3, N=3$ respectively.}
\end{figure}

From the Fig. 5, we can see that the time-average probabilities of quantum walks with $\textit{NL}=10,N=10$ is coincide to that with $\textit{NL}=3,N=10$. Also, the time-average probabilities of quantum walks with $\textit{NL}=10,N=3$ is coincide to that with $\textit{NL}=3,N=3$.  However, time-average probability of staying at outside of quantum walks with different size finally converges to $0.5$. In conclusion, $\textit{NL}$ is not as important as $N$. Furthermore, more bigger the $N$, more smaller the crossing probability.

\section{Quantum walk on Mobius strip}
\label{sec:level4}

Mobius strip is a strange and magical two-dimensional graph. This two-dimensional graph has only one surface and one boundary. Therefore, Mobius strip has some miraculous properties. For example, if the walker can cross the boundary of the Mobius strip, he can arrive a remote place instantaneously. This property is similar to the wormhole in the universe. Fig. 6 shows the Mobius strip model. $\textit{NL}$ is the length of the red line, which is half of the boundary. $N$ is the height of Mobius strip.
\begin{figure}[hbt]\label{F21}
  \center
  \includegraphics[width=10cm]{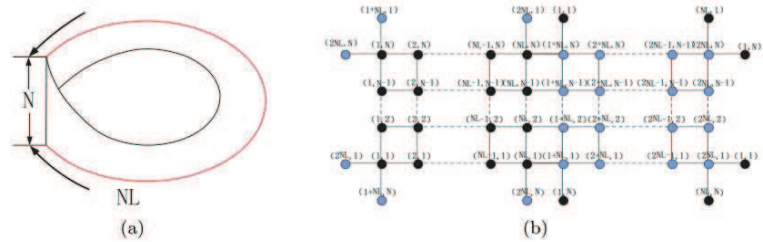}\\
  \center
  \caption{ Mobius strip model. Subgraph (b) demonstrate its corresponding lattice in two dimensions.}
\end{figure}

Because  Mobius strip has only one surface, there is no freedom of side like that in cylindrical strip. We suppose the strip is a $2\textit{NL} \times N$ lattice. The particle is living in the Hilbert space $ \mathcal{H}_p$ spanned by $|x,y\rangle$, where $x\in[0,\cdots, 2\textit{NL}-1]$, $y\in[0,\cdots, \textit{N-1}]$. The coin of the particle is in a four-dimensional Hilbert space $\mathcal{H}_c$ spanned by $\{00,01,10,11\}$. The total space for quantum walk on the Mobius strip is $\mathcal{H}=\mathcal{H}_p \otimes \mathcal{H}_c$.

One step of the quantum walk is operating a global unitary operator, denoted by $U$,
\begin{equation}
 U=S(\mathcal{I}\bigotimes C),
\end{equation}
which resembles the quantum walk on cylindrical strip. C is still the coin operation who acts on the $\mathcal{H}_c$, and S is still the walk operation. Even though there is no outside or inside for Mobius strip, the space size of the quantum walk on Mobius strip is same with that of the quantum walk on cylindrical strip with same $NL$ and $N$. That is because the $x$ direction of the particle's movement is on a longer circle.

The particle moves to left or right, up or down at one step. Therefore, S is defined by
\begin{eqnarray}
  S & = & |x+1 (mod\ \textit{2NL})\rangle\langle x|\otimes |y\rangle\langle y|\otimes|00\rangle_c \langle00|
 \nonumber\\
&& + |x\rangle\langle x|\otimes |y+1 (mod\ \textit{N})\rangle\langle y|\otimes|01\rangle_c \langle01|
\nonumber\\
&& +|x-1 (mod\ \textit{2NL})\rangle\langle x|\otimes |y\rangle\langle y|\otimes|10\rangle_c \langle10|
\nonumber\\
&& +|x\rangle\langle x|\otimes |y-1 (mod\ \textit{N})\rangle\langle y|\otimes|11\rangle_c \langle11|
\hspace{1mm}
\end{eqnarray}When $y+1\equiv 0(mod \ N)$ or $y-1\equiv \textit{N-1}(mod \ N)$, $x=x+\textit{NL}(mod\ \textit{2NL})$.

\section{properties of the quantum walk on Mobius strip}
\label{sec:level5}

Because the special topological property of Mobius strip, there are some particular properties of quantum walks on Mobius strip. In this part, we consider the properties of quantum walks on Mobius strip.

\subsection{general properties of the quantum walk on Mobius strip}
\label{sec:level51}

To begin with, we show the probability distribution of the quantum walk on Mobius strip in Fig. 7. We still chose the initial position state is $|1,1\rangle$ and the initial coin state is $[\frac{1}{2};\frac{1}{2};\frac{1}{2};\frac{1}{2}]$. $\textit{NL}$ equals 5, and $N$ equals 7.
\begin{figure}[hbt]\label{F22}
  \center
  \includegraphics[width=6cm]{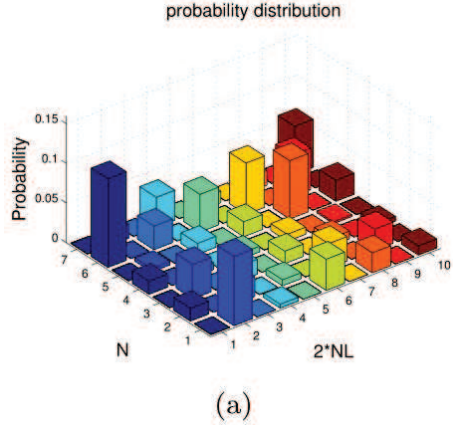}\\
  \center
  \caption{ The probability distribution of the quantum walk on Mobius strip.}
\end{figure}

For the walk on Mobius strip, no matter $\textit{NL}$ and $N$ are even or odd, when the initial position state is not entangled state there are half positions with probability zero. This is different with the quantum walk on cylindrical strip. Because for the walker on Mobius strip, the x direction of the walker's movement is on an even circle.

\subsection{Crossing the boundary}
\label{sec:level52}

Mobius strip has no another side, but when the walker on Mobius strip crossing the boundary, the walker can arrive another position which is $\sqrt{\textit{NL}^2+N^2}$ away.  So no matter for quantum walk or classical walk, the average step length is longer than $1$. It is the most magic property of Mobius strip.

Now we study the crossing probability and average step length of quantum walks and classical walks.
\begin{figure}[hbt]\label{F23}
  \center
  \includegraphics[width=10cm]{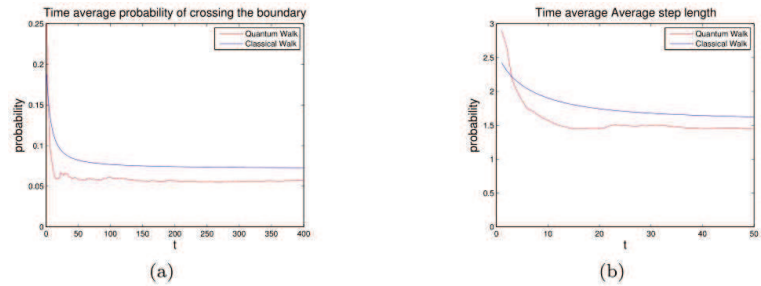}\\
  \center
  \caption{ The subgraph (a) shows the probability of crossing the boundary for quantum walks and classical walks. The subgraph (b) shows the average step length.}
\end{figure}

From the Fig. 8, we can see that the time-average crossing probability of classical walk converges to $0.0728$. Because the average step length is $S=P*\sqrt{NL^2+N^2}+(1-P)*1$, the time-average average step length converges to $1.553$.  However, the time-average crossing probability of quantum walk converges to $ 0.057$, and the time-average average step length converges $1.433$. We get the conclusion that quantum walks have lower time-average probability of crossing the boundary and lower average step length. This phenomenon  also appears in the quantum walks on cylindrical strip. It seems that quantum walks on these closed surface do not like to move compared to classical walks.

\subsection{Initial position state}
\label{sec:level53}

Then we study the effect of initial position state. For quantum walks on Mobius strip, there is one class of position states: $\alpha|x,y\rangle+\beta|x+1+2k,y\rangle$ which produce same crossing probability. $\alpha$, $\beta$ satisfy $|\alpha|^2+|\beta|^2=1$ and $k\in \mathrm{Z}$. We show their probability of crossing the boundary and average step length in Fig. 9.
\begin{figure}[hbt]\label{F24}
  \center
  \includegraphics[width=10cm]{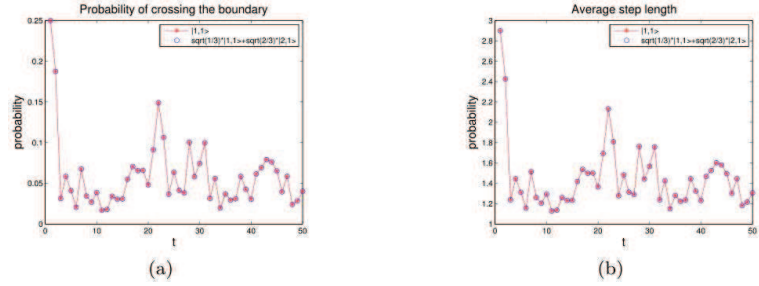}\\
  \center
  \caption{The figures show the probability of crossing the boundary and average step length for quantum walks with initial position state $|1,1\rangle$ and $\sqrt{\frac{1}{3}}|1,1\rangle+\sqrt{\frac{2}{3}}|2,1\rangle$ respectively. $\textit{NL}$=5, and $N$=7.}
\end{figure}

For this class of state as initial position state, the state is the superposition of two position state which will not coherence. Because the $x$-direction of the movement of particle is on a even circle. So, for quantum walks on Mobius strip, if initial position is is not superposed, there are half positions with probability zero. Therefore, if the initial position state is a superposition of an even position state and an odd position state,
\begin{equation}
\alpha|x,y\rangle+\beta|x+1+2k,y\rangle,
\end{equation} the finial state from these two state will not coherence.

We suppose the finial state from $|x,y\rangle$ is $\psi_{\alpha}$, and that from $|x,x+1+2k,1\rangle$ is $\psi_{\beta}$. The probability to cross the boundary is
\begin{equation}
|\alpha|^2 P_{cro}(|x,y\rangle)+|\beta|^2P_{cro}(|x+1+2k,y\rangle).
\end{equation}

Due to $|x,y\rangle$ and $|x+1+2k,y\rangle$ share same $y$ position,
\begin{equation}
P_{cro}(|x,y\rangle)=P_{cro}(|x+1+2k,y\rangle).
\end{equation}
That means the total probability to cross the boundary is $P_{cro}(|x,y\rangle)$, as long as $|\alpha|^2+|\beta|^2=1$.

\subsection{Size of Mobius strip}
\label{sec:level54}
\begin{figure}[hbt]\label{F25}
  \center
  \includegraphics[width=10cm]{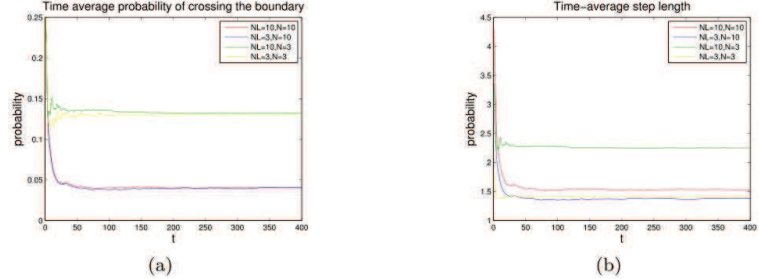}\\
  \center
  \caption{The figures show the crossing property of quantum walks on Mobius strip with $\textit{NL}=10, N=10$; $\textit{NL}=3, N=10$; $\textit{NL}=10, N=3$; $\textit{NL}=3, N=3$ respectively.}
\end{figure}
For quantum walks on Mobius strip, the size of the strip will affect the crossing probability. In Fig. 10, we can find the trend that more bigger the $N$, more small the crossing probability. It is similar to that of quantum walks on cylindrical strip. $N$ is more important than $NL$ when only considering the crossing probability. When we considering the average step length, it changes. That's because when the particle crossing the boundary, the step length is $\sqrt{NL^2+N^2}$. $NL$ is also important for average step length. More bigger the $NL$ and $N$, more bigger the step length. Therefore, concluding all above results, if the number of positions of the strip is fixed, the average step length will be bigger when $NL$ is much bigger than $N$.

\section{Conclusion}
\label{sec:level6}

In this paper, we consider single-particle discrete quantum walks on two-dimensional graphs: cylindrical strip and Mobius strip. We study the crossing property of the two kinds of quantum walks.

For quantum walks on cylindrical strip, the probability of crossing the boundary varies wildly except for one class of initial position state $\alpha|x,y,0\rangle+\beta|x,y,1\rangle$, where $|\alpha|=|\beta|=\frac{1}{2}$. For this class of  states as initial position state, the probability of crossing the boundary from outside equals to that from inside. So the probability of staying at outside or inside is 0.5 forever. Also, the height $N$ of cylindrical strip is more important than the length $NL$ when considering the crossing property of quantum walk. Furthermore, the bigger the $N$, the smaller the crossing probability.

For quantum walks on Mobius strip, there is no outside or inside. Also there is only one boundary. When crossing the boundary, the walker can get into another remote position. So the average step length is bigger than $1$. For quantum walks with a class of states $\alpha|x,y\rangle+\beta|x+1+2k,y\rangle$ as initial position state, the crossing probability and average step length don't change. Furthermore, the smaller the $N$, the bigger the crossing probability. At the same time, the bigger the $NL$ and $N$, the bigger the average step length. Therefore, if we want to get a bigger average step length, $NL$ should be much bigger than $N$.

\section*{Acknowledgements}

This work is supported by NSFC (Grant Nos. 61300181, 61272057, 61202434, 61170270, 61100203, 61121061), Beijing Natural Science Foundation (Grant No. 4122054), Beijing Higher Education Young Elite Teacher Project, BUPT Excellent Ph.D. Students Foundation(Grant Nos. CX201325, CX201326), China Scholarship Council(Grant Nos. 201306470046).

\vspace*{2mm}

%%%% ²Î¿¼ÎÄÏ×ÅÅ°æ¸ñʽ£º

 %% ½áÊøÖÐÎÄ¡¢ÈÕÎÄ¡¢º«ÎÄʹÓû·¾³
\end{document}